\documentclass[letterpaper,10pt,conference]{ieeeconf}
\IEEEoverridecommandlockouts
\usepackage{cite}
\usepackage{amsmath,amssymb,amsfonts}
\allowdisplaybreaks[4]
\usepackage{algorithm}
\usepackage{algpseudocode}
\algrenewcommand\algorithmicrequire{\textbf{Input:}}
\algrenewcommand\algorithmicensure{\textbf{Output:}}
\usepackage{graphicx}

\usepackage{textcomp}
\usepackage{microtype}
\usepackage{url}
\usepackage{booktabs}
\usepackage{xcolor}
\usepackage{tikz}
\usepackage{mathtools}
\usetikzlibrary{backgrounds}
\usetikzlibrary{arrows.meta,positioning,calc,shapes.geometric,fit}

\def\BibTeX{{\rm B\kern-.05em{\sc i\kern-.025em b}\kern-.08em
    T\kern-.1667em\lower.7ex\hbox{E}\kern-.125emX}}

\makeatletter
\let\NAT@parse\undefined
\makeatother
\usepackage[hidelinks]{hyperref}
\makeatletter
\long\def\@makecaption#1#2{%
  \par\addvspace{0.5\baselineskip}%
  \begingroup\footnotesize
  \def\tempa{table}%
  \ifx\@captype\tempa
    \centering #1\par\vspace{2pt}\textsc{#2}\par\vspace{4pt}%
  \else
    \setbox\@tempboxa\hbox{#1. #2}%
    \ifdim\wd\@tempboxa>\hsize #1. #2\par
    \else\hbox to\hsize{\hfil\box\@tempboxa\hfil}\fi
  \fi\endgroup}
\makeatother

\title{
\LARGE \bfseries Reducing Combinatorial Redundancy in Mixed-Integer MPC via
Ranking-Based Feasible-Set Restriction for Reconfigurable Batteries
}

\author{Albert \v{S}kegro, Torsten Wik, and Changfu Zou%
\thanks{This work was supported by the Swedish Research Council (Grant No.~2023--04314) and the European Union's Horizon Europe program through the Marie Sk\l{}odowska--Curie Actions (Grant No.~101131278). The computations were enabled by resources provided by Chalmers e-Commons at Chalmers.}%
\thanks{\sloppy All authors are with the Department of Electrical Engineering, Chalmers University of Technology, Gothenburg, Sweden. (e-mails: skegro@chalmers.se, torsten.wik@chalmers.se, changfu.zou@chalmers.se).}
}

\begin{document}
\bstctlcite{BSTcontrol}

\maketitle
\thispagestyle{empty}
\pagestyle{empty}
\raggedbottom
\setlength{\abovedisplayskip}{3pt plus 1pt minus 1pt}
\setlength{\belowdisplayskip}{3pt plus 1pt minus 1pt}
\setlength{\abovedisplayshortskip}{3pt plus 1pt minus 1pt}
\setlength{\belowdisplayshortskip}{3pt plus 1pt minus 1pt}
\setlength{\textfloatsep}{8pt plus 2pt minus 2pt}
\setlength{\intextsep}{8pt plus 2pt minus 2pt}
\setlength{\floatsep}{8pt plus 2pt minus 2pt}

\begin{abstract}
When mixed-integer model predictive control must select which of several similar units to activate, many same-size selections can yield closely spaced predicted outcomes, and branch-and-bound may fail to certify optimality within one sampling period. We propose a ranking-based feasible-set restriction for this class of decisions: the units are ordered by a suitability score computed at each sampling instant, and a monotone constraint permits only the top-ranked units to be active. The optimizer then chooses only how many units to activate, so the admissible selections per prediction stage grow linearly instead of combinatorially, while the model, cost function, and remaining constraints are retained. We instantiate the method for reconfigurable battery packs in which each series-connected cell can be bypassed, scoring cells with fixed weights on normalized state of health, state of charge, and voltage headroom. In a 20-cell simulation over a dynamic drive cycle with a 1 s limit per online solve, all active online restricted solves terminate at the root node with certified optimality, compared with 77.83\% certification for unrestricted subset selection, and the 95th-percentile solver time is lower by a factor of 13.4. The cycle-mean cell-to-cell standard deviation in state of charge falls 8.7-fold without power curtailment, at the expense of 3.45 times as much cell switching. Solving both problems from identical states along two trajectories shows that the restriction raises the optimal cost by 0.26\% and 0.14\% on average at states where both solves are certified and uncurtailed.
\end{abstract}

\section{Introduction}
\label{sec:introduction}

Mixed-integer model predictive control (MI-MPC) computes discrete operating decisions together with continuous control inputs under
hybrid dynamics, logical relations, and physical bounds~\cite{rawlings2020model,bemporad1999control,richards2005mixed}. Online operation places a deadline on this optimization at every control update. Finding an admissible input and proving its optimality require different amounts of computation: a branch-and-bound solver finds incumbents during the search and certifies optimality when the gap meets the prescribed tolerance~\cite{marcucci2020warm,hespanhol2019structure}. Variation in the time needed to close that gap makes a fixed sampling period difficult to meet.

A recurring discrete decision in MI-MPC is the selection of which of several similar units to activate. Once the number of active units is
fixed, many selections of that size remain. Small symmetric integer programs are extremely difficult for branch-and-cut codes that ignore
the symmetry~\cite{margot2009symmetry}, and in unit commitment, aggregating identical generators, which removes symmetric and equal-cost alternatives, reduces branch-and-cut node counts by an order of magnitude or more on instances with many such generators~\cite{knueven2018}. When the units are similar, many same-size selections have closely spaced predicted costs, and certifying optimality among them within one sampling period becomes difficult.

For exact symmetries, symmetry-handling techniques remove redundant solutions without changing the optimum~\cite{margot2009symmetry}. When the units are heterogeneous, however, same-size selections have different physical effects; the redundancy is approximate, and exact symmetry handling does not apply. Methods that retain mixed-integer optimization instead act on the search procedure: warm starts and structured branch-and-bound reuse information and exploit the optimal-control problem structure~\cite{marcucci2020warm,hespanhol2019structure}, and distributionally robust lifted feedback policies treat uncertainty~\cite{ma2025datadriven}. These methods leave the set of admissible same-size selections unchanged.

Reconfigurable battery packs (RBPs) provide a representative instance of this decision class. In a series-connected RBP, bypass switches determine which cells carry pack current. This control over cell use supports balancing, fault tolerance, and capacity utilization~\cite{han2020next,muhammad2019reconfigurable,ci2016reconfigurable}; lifetime and cost benefits depend on pack design and operating conditions~\cite{vskegro2026system}. When cells have similar state of charge (SOC), temperature, resistance, and aging condition, different subsets yield closely spaced terminal voltages and predicted states.

Predictive battery controllers treat switching in different ways. Modular and converter-integrated packs admit convex receding-horizon power allocation~\cite{farakhor2022novel,farakhor2023scalable}, and for small bypass-switched packs, nonlinear model predictive control (MPC) prioritizes feasible control over a global-optimality certificate~\cite{mondoha2021nonlinear}. In a commercial reconfigurable system, MI-MPC outperforms rule-based control but requires at least two orders of magnitude more computation~\cite{pinter2025comparative}. Control allocation with barrier functions~\cite{ebrahimi2026safe} and learning-based policies~\cite{wei2026reliable,liu2026learning,irshayyid2026realtime} replace online mixed-integer optimization with alternative control laws. In learning-based control of series packs, the size of the action space equals the number of engaged-cell subsets, and the training effort grows steeply with pack size; preselecting the cells with the lowest state of health (SOH) as bypass candidates is proposed as a future extension~\cite{liu2026learning}. Switch-dependent models and feasible-search-space construction exclude invalid or duplicate topologies~\cite{geng2025fundamental,geng2026unified}, but many physically admissible subsets with the same cell count remain.

We propose a ranking-based feasible-set restriction that removes same-size selections from the MI-MPC decision set. At each sampling instant, the units are ranked by a suitability score, and linear constraints allow only the top-ranked units to be active, so the optimizer, in selecting the active subset, chooses only how many units to activate. This reduces the admissible selections per prediction stage from combinatorial to linear in the number of units, leaves the model, cost function, and other constraints unchanged, and requires no offline training or solver modification. We apply the restriction to series-connected RBPs, ranking cells by SOH, SOC, and voltage headroom, and evaluate its closed-loop performance and objective cost against full subset selection.

\section{Control Model and MI-MPC Formulation}
\label{sec:modelAndProblemFormulation}

The series-connected pack contains $N$ cells, with cell index $i\in\mathcal C:=\{1,\ldots,N\}$. The sampling instant, sample time, and prediction length are $k$, $\Delta t$, and $N_p$, respectively; $\ell\in\mathcal H_k:=\{k,\ldots,k+N_p-1\}$ indexes prediction stages. The binary series-switch state $S_i$ indicates engagement when $S_i=1$ and bypass when $S_i=0$. The complementary bypass-switch state is $S_i^{\prime}=1-S_i$. The superscript $\top$ denotes transpose. The vector $S=(S_1,\ldots,S_N)^\top$ collects these states, and $m=\sum_{i\in\mathcal C}S_i$ gives the engagement cardinality.

\subsection{Control-Oriented Pack Model}
\label{subsec:electhermBattModel}
Each cell follows a resistor--capacitor (RC) electrical model coupled to a lumped heat balance~\cite{vskegro2023analysis,ouyang2025mathematical,allafi_lumped_2018}. We collect the SOC of cell $i$, its RC-branch current $i_{\mathrm{RC},i}$, and temperature $T_i$ in $x_i=(\mathrm{SOC}_i,i_{\mathrm{RC},i},T_i)^\top$ and stack the cell states as $x=(x_1^\top,\ldots,x_N^\top)^\top$. The ratio of cell capacity $Q_i$ to nominal capacity $Q^{\mathrm{nom}}$ defines $\mathrm{SOH}_i=Q_i/Q^{\mathrm{nom}}$. The cell capacities $Q_i$ and health values $\mathrm{SOH}_i$ are held fixed over each prediction horizon.

We define current as positive during discharge and negative during charge. The symbols $i_{\mathrm{pack}}$ and $i_i$ denote pack and cell current; $v_{\mathrm{pack}}$, $v_i$, and $p_{\mathrm{pack}}$ denote pack voltage, cell voltage, and pack power. The open-circuit voltage (OCV) map is $v_{\mathrm{OC},i}(\cdot)$. With ohmic resistance $R_0$, RC-branch resistance $R_1$, and switch resistance $r_s$, the electrical relations are
\begin{subequations}\label{eq:packModel}
\begin{align}
i_i&=S_i i_{\mathrm{pack}},\label{eq:cellCurrent}\\
v_i&=v_{\mathrm{OC},i}(\mathrm{SOC}_i)-R_0i_i-R_1i_{\mathrm{RC},i},\label{eq:cellVoltage}\\
v_{\mathrm{pack}}&=\sum_{i\in\mathcal C}S_iv_i-Nr_si_{\mathrm{pack}},\label{eq:packVoltage}\\
p_{\mathrm{pack}}&=i_{\mathrm{pack}}v_{\mathrm{pack}}.\label{eq:packPower}
\end{align}
\end{subequations}
Let $C_1$ be the RC-branch capacitance, $C_{\mathrm{th}}$ the cell-unit heat capacity, $h_{\mathrm{amb}}$ the thermal conductance to the surroundings, and $T_{\mathrm{amb}}$ the ambient temperature. Define the discrete-time coefficients
\begin{equation}
\begin{aligned}
a_1&=\exp[-\Delta t/(R_1C_1)], & b_1&=1-a_1,\\
a_T&=\exp[-h_{\mathrm{amb}}\Delta t/C_{\mathrm{th}}], & b_T&=(1-a_T)/h_{\mathrm{amb}}.
\end{aligned}
\end{equation}
With $Q_i$ expressed in ampere-hours, the cell dynamics are
\begin{subequations}\label{eq:cellDynamicsExpanded}
\begin{align}
\mathrm{SOC}_i(\ell+1)&=\mathrm{SOC}_i(\ell)-\frac{\Delta t}{3600Q_i}i_i(\ell),\\
i_{\mathrm{RC},i}(\ell+1)&=a_1i_{\mathrm{RC},i}(\ell)+b_1i_i(\ell),\\
T_i(\ell+1)&=T_{\mathrm{amb}}+a_T[T_i(\ell)-T_{\mathrm{amb}}]\notag\\
&\quad+b_T\dot Q_i^{\mathrm{gen}}(\ell),
\end{align}
\end{subequations}
where the cell-unit heat-generation rate $\dot Q_i^{\mathrm{gen}}$ includes cell and switch losses:
\begin{equation}\label{eq:heatGeneration}
\dot Q_i^{\mathrm{gen}}=R_0i_i^2+R_1i_{\mathrm{RC},i}^2+r_si_{\mathrm{pack}}^2.
\end{equation}
Pack current passes through one switch per cell unit, so each unit contributes switch heat whether its cell is engaged or bypassed. The state-transition map $f$ collects~\eqref{eq:cellDynamicsExpanded}. With input $u=(S^\top,i_{\mathrm{pack}},\lambda)^\top$ and current-equivalent curtailment slack $\lambda$, the pack dynamics are $x(\ell+1)=f(x(\ell),u(\ell))$.

\begin{figure*}[t]
\centering
\resizebox{\textwidth}{!}{\input{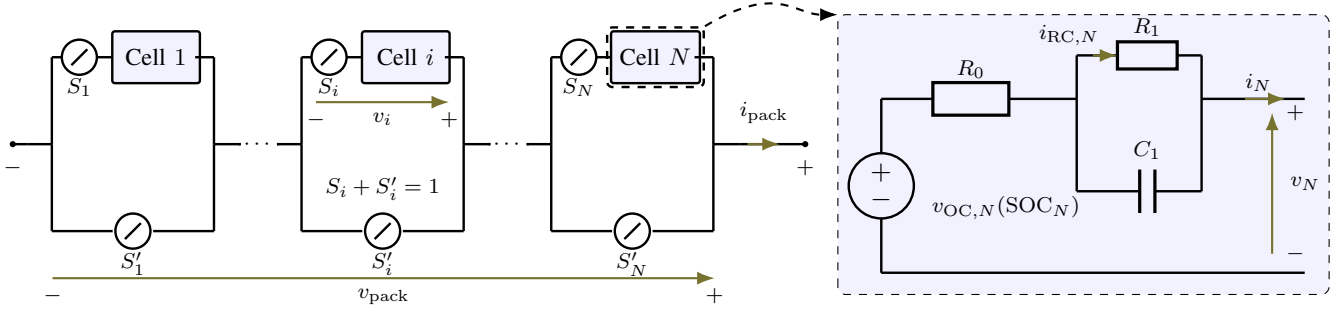}}
\caption{Series-connected reconfigurable battery pack with a series switch $S_i$, complementary bypass switch $S_i^{\prime}$, and a first-order RC equivalent circuit for each cell.}
\label{fig:pack_topology}
\end{figure*}

\subsection{Control Objective and Constraints}
\label{subsec:MI-MPCprobFor}
The requested pack power is $p_{\mathrm{req}}$. Its direction indicator is $d_p=1$ for a nonnegative request and $d_p=-1$ for a negative request.\label{eq:powerDirectionText}
The charge and discharge current limits satisfy $i_{\mathrm{pack}}^{\min}<0<i_{\mathrm{pack}}^{\max}$. For a scalar, $|\cdot|$ denotes absolute value. The current scale $i_{\mathrm{scale}}$ equals $i_{\mathrm{pack}}^{\max}$ for nonnegative requests and $|i_{\mathrm{pack}}^{\min}|$ otherwise. The admissible current set is $\mathcal I=[0,i_{\mathrm{pack}}^{\max}]$ during discharge, $\mathcal I=[i_{\mathrm{pack}}^{\min},0]$ during charge, and $\mathcal I=\{0\}$ at zero request. The slack expresses curtailed power through
\begin{equation}
\lambda=\frac{p_{\mathrm{req}}-p_{\mathrm{pack}}}{d_pv_{\mathrm{pack}}},\qquad
0\leq\lambda\leq\frac{\delta_P|p_{\mathrm{req}}|}{v_{\mathrm{pack}}^{\min}},
\label{eq:lambdaDef}
\end{equation}
where $\delta_P\in(0,1)$ is the curtailment coefficient and $v_{\mathrm{pack}}^{\min}>0$ the pack-voltage lower limit. Zero request imposes zero current and slack.

Lower and upper bounds carry superscripts $\min$ and $\max$. For cell SOC, temperature, and voltage, these bounds are $\mathrm{SOC}_{\mathrm{cell}}^{\min,\max}$, $T_{\mathrm{cell}}^{\min,\max}$, and $v_{\mathrm{cell}}^{\min,\max}$; the upper pack-voltage limit is $v_{\mathrm{pack}}^{\max}$. The admissible state set $\mathcal X$ contains states whose SOC and temperature satisfy these bounds; the state constraint is $x(\ell+1)\in\mathcal X$. Define the mean SOC as $\bar{\mathrm{SOC}}=\frac{1}{N}\sum_{i\in\mathcal C}\mathrm{SOC}_i$. The terminal SOC-spread bound $\Delta\mathrm{SOC}^{\max}(k)$ depends on the lowest cell SOC, $z_{\min}(k)=\min_{i\in\mathcal C}\mathrm{SOC}_i(k)$. Let $z_1<z_2$ be the SOC thresholds delimiting a transition band and $\Delta_{\mathrm{low}}\leq\Delta_{\mathrm{high}}$ the minimum and maximum target spreads. The interpolation factor $\eta(k)$ and target spread $\Delta\mathrm{SOC}_{\mathrm{tar}}^{\max}(k)$ are
\begin{subequations}\label{eq:spreadTarget}
\begin{align}
\eta(k)&=\min\left\{1,\max\left\{0,\frac{z_{\min}(k)-z_1}{z_2-z_1}\right\}\right\},\\
\Delta\mathrm{SOC}_{\mathrm{tar}}^{\max}(k)&=\Delta_{\mathrm{low}}\notag\\
&\quad+(\Delta_{\mathrm{high}}-\Delta_{\mathrm{low}})\eta(k).
\end{align}
\end{subequations}
The maximum rate of decrease of the terminal spread bound is $s_\Delta>0$, in SOC fraction per second. The update
\begin{equation}\label{eq:spreadUpdate}
\begin{aligned}
\Delta\mathrm{SOC}^{\max}(k)=\max\bigl\{&\Delta\mathrm{SOC}_{\mathrm{tar}}^{\max}(k),\\
&\Delta\mathrm{SOC}^{\max}(k-1)-s_\Delta\Delta t\bigr\}
\end{aligned}
\end{equation}
limits tightening to $s_\Delta\Delta t$ per sample while allowing immediate relaxation. The previous bound is initialized to $\Delta_{\mathrm{high}}$ before the first update.

With end-of-life health threshold $\mathrm{SOH}^{\mathrm{EOL}}$, the dimensionless health, balancing, and curtailment costs $J_{\mathrm{SOH}}$, $J_{\mathrm{SOC}}$, and $J_\lambda$ are
\begin{subequations}\label{eq:stageCosts}
\begin{align}
J_{\mathrm{SOH}}(\ell)&=\frac{1}{N}\sum_{i\in\mathcal C}\frac{\mathrm{SOH}^{\mathrm{EOL}}}{\mathrm{SOH}_i}S_i(\ell),\label{eq:JSOH}\\
J_{\mathrm{SOC}}(\ell)&=\frac{1}{N}\sum_{i\in\mathcal C}\left(\frac{\mathrm{SOC}_i(\ell+1)-\bar{\mathrm{SOC}}(\ell+1)}{\Delta\mathrm{SOC}^{\max}(k)}\right)^2,\label{eq:JSOC}\\
J_\lambda(\ell)&=\lambda(\ell)/i_{\mathrm{scale}}(\ell).\label{eq:Jlambda}
\end{align}
\end{subequations}
Nonnegative weights $w_{\mathrm{SOH}}$, $w_{\mathrm{SOC}}$, and $w_\lambda$ combine these costs into the horizon objective $J(k)$:
\begin{align}
J(k)=\frac{1}{N_p}\sum_{\ell\in\mathcal H_k}\bigl[&w_{\mathrm{SOH}}J_{\mathrm{SOH}}(\ell)+w_{\mathrm{SOC}}J_{\mathrm{SOC}}(\ell)\notag\\
&+w_\lambda J_\lambda(\ell)\bigr].\label{eq:ContrOBJ}
\end{align}
The stage-dependent minimum engagement count is $m_{\min}(\ell)$, and the decision set is $\mathcal U_k=\{u(\ell),x(\ell+1):\ell\in\mathcal H_k\}$. Full subset selection solves
\begin{subequations}\label{eq:unrestrictedMIMPC}
\begin{align}
&\min_{\mathcal U_k}J(k)\label{eq:unrestrictedMIMPC_obj}\\
\text{s.t. }&x(\ell+1)=f(x(\ell),u(\ell)),\quad x(k)\text{ given},\\
&S_i(\ell)\in\{0,1\},\quad m(\ell)\geq m_{\min}(\ell),\label{eq:commonCardinalityFloor}\\
&i_{\mathrm{pack}}(\ell)\in\mathcal I(\ell),\quad x(\ell+1)\in\mathcal X,\\
&v_i(\ell)\in[v_{\mathrm{cell}}^{\min},v_{\mathrm{cell}}^{\max}],\\
&v_{\mathrm{pack}}(\ell)\in[v_{\mathrm{pack}}^{\min},v_{\mathrm{pack}}^{\max}],\\
&\max_{i\in\mathcal C}\mathrm{SOC}_i(k+N_p)-\min_{i\in\mathcal C}\mathrm{SOC}_i(k+N_p)\notag\\
&\hspace{35mm}\leq\Delta\mathrm{SOC}^{\max}(k),\label{eq:termSOCSpreadConstraint}
\end{align}
\end{subequations}
for every $i\in\mathcal C$ and $\ell\in\mathcal H_k$, together with the electrical relations~\eqref{eq:packModel} and curtailment relation and bounds~\eqref{eq:lambdaDef}.

\subsection{Mixed-Integer Convex Implementation}
\label{subsec:micp}
The implementation approximates~\eqref{eq:unrestrictedMIMPC} by a mixed-integer convex program (MICP). Binary--continuous products are expressed through auxiliary variables with linear constraints derived from variable bounds~\cite{bemporad1999control,williams2013model}. Each prediction horizon uses one fixed local affine segment of the OCV curve. For reciprocal pack voltage, let $M_v$ denote the number of positive, increasing voltage breakpoints $V_q$, indexed by $q\in\{1,\ldots,M_v\}$, spanning the admissible pack-voltage interval. Interpolation weights $\alpha_q(\ell)$ link pack voltage and its reciprocal approximation $r_{\mathrm{pack}}(\ell)$ through
\begin{equation}
\begin{alignedat}{2}
v_{\mathrm{pack}}(\ell)&=\sum_{q=1}^{M_v}V_q\alpha_q(\ell),&\quad r_{\mathrm{pack}}(\ell)&=\sum_{q=1}^{M_v}\frac{\alpha_q(\ell)}{V_q},\\
\sum_{q=1}^{M_v}\alpha_q(\ell)&=1,& \alpha_q(\ell)&\geq0.
\end{alignedat}
\end{equation}
The weights $\{\alpha_q(\ell)\}_{q=1}^{M_v}$ form a special ordered set of type two: at most two weights are positive, and their indices correspond to adjacent voltage breakpoints $V_q$. A conic epigraph represents squared pack current for thermal prediction, which retains ohmic and switch losses; the simulation plant also retains RC polarization losses. To state the resulting optimization problem, let $\mathcal V_k$ collect the variables in $\mathcal U_k$ and the auxiliary variables, and let $\mathcal F_k^{\mathrm{MICP}}$ denote their feasible set. The full subset-selection MICP is
\begin{equation}\label{eq:fullMICP}
\min_{\mathcal V_k}\ J(k)\qquad\text{s.t.}\quad\mathcal V_k\in\mathcal F_k^{\mathrm{MICP}}.
\end{equation}
Its optimal objective is $J_{\mathrm{full}}^\star(k)$, where $\star$ denotes an optimum.

\section{Combinatorial Structure of Cell Engagement}
\label{sec:sourceRedundancy}

At each stage $\ell$, the full subset-selection problem~\eqref{eq:fullMICP} chooses a binary engagement vector $S(\ell)\in\{0,1\}^N$. Its cardinality $m(\ell)$ determines how many cell voltages contribute to the pack voltage and affects the current needed to meet the power request. Grouping candidates by cardinality makes the size of the binary search domain explicit. For a fixed count $m_{\mathrm{fix}}\in\{0,\ldots,N\}$, the same-cardinality class $\mathcal S_{m_{\mathrm{fix}}}$ is
\begin{align}
\mathcal S_{m_{\mathrm{fix}}}&=\{S\in\{0,1\}^N:\textstyle\sum_{i\in\mathcal C}S_i=m_{\mathrm{fix}}\},\notag\\
|\mathcal S_{m_{\mathrm{fix}}}|&=\binom{N}{m_{\mathrm{fix}}}.\label{eq:samecard_class}
\end{align}
Here, $|\cdot|$ denotes set cardinality and $\binom{N}{m_{\mathrm{fix}}}$ counts subsets of that size. Choosing among same-cardinality subsets creates the combinatorial search burden.

Idealized voltage and power capability screen low-cardinality candidates. Following the series-cell voltage screen in~\cite{geng2025fundamental}, the voltage-based minimum count $m_{\min}^{(V)}$ is
\begin{equation}
m_{\min}^{(V)}=\left\lceil\frac{v_{\mathrm{pack}}^{\min}}{v_{\mathrm{cell}}^{\max}}\right\rceil,\label{eq:m_min_V}
\end{equation}
where $\lceil\cdot\rceil$ denotes rounding upward to an integer. The direction-dependent hardware current ceiling $i_{\mathrm{hw}}(\ell)=i_{\mathrm{scale}}(\ell)$ determines the full-request power-based count $m_{\min}^{(P)}(\ell)$:
\begin{equation}
m_{\min}^{(P)}(\ell)=\left\lceil\frac{|p_{\mathrm{req}}(\ell)|}{v_{\mathrm{cell}}^{\max}i_{\mathrm{hw}}(\ell)}\right\rceil.\label{eq:m_min_P}
\end{equation}
These counts determine the cardinality floor
\begin{equation}
m_{\min}(\ell)=\min\{\max\{m_{\min}^{(V)},m_{\min}^{(P)}(\ell)\},N\}.\label{eq:m_min}
\end{equation}
The cardinality floor uses the full power request, including when curtailment is allowed. The dynamics and constraints in~\eqref{eq:unrestrictedMIMPC} determine physical feasibility.

The resulting binary candidate domain $\mathcal S_{\mathrm{cand}}(\ell)$ and its cardinality satisfy
\begin{align}
\mathcal S_{\mathrm{cand}}(\ell)&=\{S\in\{0,1\}^N:\textstyle\sum_{i\in\mathcal C}S_i\geq m_{\min}(\ell)\},\notag\\
|\mathcal S_{\mathrm{cand}}(\ell)|&=\sum_{p=m_{\min}(\ell)}^N\binom Np,\label{eq:candidateDomain}
\end{align}
where $p$ indexes the integer engagement count. The domain contains $N-m_{\min}(\ell)+1$ possible cardinalities, with multiple candidate subsets at each intermediate cardinality.

\section{Ranking-Based Feasible-Set Restriction}
\label{sec:restriction}

We rank the cells by their current SOH, SOC, and voltage headroom. An engagement prefix contains the first $m$ cells in this ranking; all remaining cells are bypassed. Thus, choosing $m$ also determines which cells are engaged.

\subsection{Three-Component Cell Ranking}
Current cell states determine the ranking at sampling instant $k$. For a cell-indexed quantity $\xi_i$, let $\bar\xi=\frac{1}{N}\sum_{i\in\mathcal C}\xi_i$ denote its cross-cell mean and $\epsilon$ a positive normalization tolerance. The range-normalization operator $\mathcal R$, with cell index $j\in\mathcal C$, is
\begin{equation}
\mathcal R(\xi_i)=\begin{cases}
\displaystyle\frac{\xi_i-\bar\xi}{\max_j\xi_j-\min_j\xi_j},&\max_j\xi_j-\min_j\xi_j>\epsilon,\\
0,&\text{otherwise},
\end{cases}\label{eq:normalization}
\end{equation}
The operator normalizes each component across the cells and assigns zero preference when its range falls within the tolerance.

The normalized health and charge preferences of cell $i$ are $\rho_i^{\mathrm{SOH}}$ and $\rho_i^{\mathrm{SOC}}$, respectively:
\begin{align}
\rho_i^{\mathrm{SOH}}(k)&=\mathcal R(-1/\mathrm{SOH}_i);\\
\rho_i^{\mathrm{SOC}}(k)&=\mathcal R(d_p(k)\mathrm{SOC}_i(k)).
\end{align}
These terms favor healthier cells and, through the requested-power direction $d_p(k)$, higher SOC during discharge or lower SOC during charge.

The representative engagement count is $m_{\mathrm{ref}}(k)=\max\{m_{\min}^{(V)},m_{\min}^{(P)}(k),1\}$. The representative OCV $v_{\mathrm{OC,ref}}(k)$ equals the minimum cell OCV during discharge and the maximum during charge. Voltage-headroom estimation uses the representative current magnitude $i_{\mathrm{exp}}(k)$:
\begin{equation}
i_{\mathrm{exp}}(k)=\min\left\{\frac{|p_{\mathrm{req}}(k)|}{v_{\mathrm{OC,ref}}(k)m_{\mathrm{ref}}(k)},i_{\mathrm{hw}}(k)\right\}.\label{eq:iexp_rank}
\end{equation}
The terminal-voltage estimate $\hat v_i(k)$ is
\begin{equation}
\hat v_i(k)=v_{\mathrm{OC},i}(k)-d_p(k)R_0i_{\mathrm{exp}}(k)-R_1i_{\mathrm{RC},i}(k),\label{eq:vhat_rank}
\end{equation}
where $v_{\mathrm{OC},i}(k)$ abbreviates $v_{\mathrm{OC},i}(\mathrm{SOC}_i(k))$. The clipped estimate is $\hat v_i^{\mathrm{clip}}(k)=\min\{v_{\mathrm{cell}}^{\max},\max\{v_{\mathrm{cell}}^{\min},\hat v_i(k)\}\}$. It gives the voltage headroom $h_i(k)$ and normalized voltage preference $\rho_i^v(k)$:
\begin{align}
h_i(k)&=\begin{cases}
\hat v_i^{\mathrm{clip}}(k)-v_{\mathrm{cell}}^{\min},&p_{\mathrm{req}}(k)\geq0,\\
v_{\mathrm{cell}}^{\max}-\hat v_i^{\mathrm{clip}}(k),&p_{\mathrm{req}}(k)<0,
\end{cases}\\
\rho_i^v(k)&=\mathcal R(h_i(k)).
\end{align}
This term favors greater voltage headroom. Zero request uses the discharge convention and $i_{\mathrm{exp}}=0$.

\paragraph{Composite Score and Cell Ordering}
Fixed nonnegative coefficients $\gamma_v$, $\gamma_{\mathrm{SOC}}$, and $\gamma_{\mathrm{SOH}}$ weight voltage headroom, SOC, and SOH, respectively, in the composite suitability score $\rho_i(k)$:
\begin{equation}
\rho_i(k)=\gamma_v\rho_i^v(k)+\gamma_{\mathrm{SOC}}\rho_i^{\mathrm{SOC}}(k)+\gamma_{\mathrm{SOH}}\rho_i^{\mathrm{SOH}}(k).
\label{eq:rankingScore}
\end{equation}
Higher scores indicate greater suitability for engagement. The vector $\boldsymbol\rho(k)$ collects the scores:
\begin{equation}
\boldsymbol\rho(k)=\big(\rho_1(k),\ldots,\rho_N(k)\big)^\top.
\label{eq:rho_vector}
\end{equation}
The operator $\operatorname{argsort}$ returns the permutation that sorts its argument in nondecreasing order. The descending-score permutation $\pi_k$ therefore satisfies
\begin{equation}
\pi_k=\operatorname{argsort}\big(-\boldsymbol\rho(k)\big),
\label{eq:argsort}
\end{equation}
where $\pi_k:\{1,\ldots,N\}\to\{1,\ldots,N\}$ maps the rank position $j\in\{1,\ldots,N\}$ to the physical index of the cell ranked $j^{\mathrm{th}}$. The resulting order satisfies
\begin{equation}
\rho_{\pi_k(1)}(k)\geq\rho_{\pi_k(2)}(k)\geq\cdots\geq\rho_{\pi_k(N)}(k).
\label{eq:score_order}
\end{equation}
For equal scores, the cell with the lower physical index comes first. We apply $\pi_k$ to every cell-indexed variable, parameter, and constraint and keep this order throughout the horizon.

\subsection{Monotone Prefix Constraint}
The ranked engagement variables $S^{\mathrm{rank}}_j(\ell)=S_{\pi_k(j)}(\ell)$ satisfy the monotone constraint
\begin{equation}
S^{\mathrm{rank}}_j(\ell)\geq S^{\mathrm{rank}}_{j+1}(\ell),\quad j=1,\ldots,N-1,\quad\ell\in\mathcal H_k.
\label{eq:monotone}
\end{equation}
This constraint engages the first $m(\ell)$ ranked cells and bypasses the rest, with $m(\ell)=\sum_{j=1}^{N}S^{\mathrm{rank}}_j(\ell)\in\{m_{\min}(\ell),\ldots,N\}$. The $N-1$ additional linear inequalities per stage reduce the candidate count from $\sum_{p=m_{\min}(\ell)}^N\binom Np$ to $N-m_{\min}(\ell)+1$. Adding these inequalities to~\eqref{eq:fullMICP} gives the restricted problem
\begin{equation}\label{eq:restrictedMICP}
\begin{aligned}
\min_{\mathcal V_k}\quad &J(k)\\
\text{s.t.}\quad &\mathcal V_k\in\mathcal F_k^{\mathrm{MICP}},\\
&S^{\mathrm{rank}}_j(\ell)\geq S^{\mathrm{rank}}_{j+1}(\ell),\quad j=1,\ldots,N-1,\\
&\hspace{44mm}\ell\in\mathcal H_k.
\end{aligned}
\end{equation}
The optimal objective is $J_{\mathrm{pref}}^\star(k)$. The objective, prediction equations, and physical constraints are those of~\eqref{eq:fullMICP}. For example, a stage with $m(\ell)=5$ engages the five highest-ranked cells; a later stage can choose a different count and current. The cell ranking used for these stagewise choices is the one computed at sampling instant $k$.

\subsection{Online Implementation}
Algorithm~\ref{alg:onlineRankedPrefixMI-MPC} gives the online procedure. With $\mathcal O$ denoting asymptotic complexity, evaluating the scores takes $\mathcal O(N)$ operations, sorting takes $\mathcal O(N\log N)$ time, and storing the scores and cell permutation requires $\mathcal O(N)$ memory. A standard MICP solver handles the resulting optimization problem.

\begin{algorithm}[!t]
\caption{Online ranked-prefix MI-MPC}
\label{alg:onlineRankedPrefixMI-MPC}
\begin{algorithmic}[1]
\Require Cell states and SOH; $\{p_{\mathrm{req}}(\ell)\}_{\ell\in\mathcal H_k}$; previous bound $\Delta\mathrm{SOC}^{\max}(k-1)$.
\Ensure Engagement $S(k)$ and pack current $i_{\mathrm{pack}}(k)$.
\State Update the cardinality floor~\eqref{eq:m_min_V}--\eqref{eq:m_min}; update the terminal bound using~\eqref{eq:spreadUpdate}.
\State Evaluate the scores~\eqref{eq:normalization}--\eqref{eq:rankingScore}.
\State Obtain the ordering $\pi_k$ from~\eqref{eq:argsort}.
\State Reorder all cell-indexed quantities according to $\pi_k$ and impose~\eqref{eq:monotone}.
\State Solve the restricted MICP~\eqref{eq:restrictedMICP} within the time limit.
\State Apply the first action of the returned feasible solution in physical cell order; if no feasible solution is returned, set $S_i=0$ for all cells and $i_{\mathrm{pack}}=\lambda=0$.
\end{algorithmic}
\end{algorithm}

\section{Numerical Evaluation Setup}
\label{sec:evalMethodology}
We consider a 20-cell lithium iron phosphate pack with the configuration in Figure~\ref{fig:pack_topology}, evaluated over the Worldwide Harmonized Light Vehicles Test Cycle (WLTC). We refer to the ranked-prefix controller solving~\eqref{eq:restrictedMICP} as the proposed controller and the controller solving the full subset-selection MICP~\eqref{eq:fullMICP} as the baseline controller.

\subsection{Simulation and Solver Settings}
The electrothermal equations~\eqref{eq:packModel}--\eqref{eq:heatGeneration} propagate the battery-pack states during closed-loop simulation. Both controllers use the mixed-integer convex model in~\eqref{eq:fullMICP}, with identical initial conditions, requested power, objective weights, physical limits, and solver settings. MATLAB and YALMIP construct the optimization problems, which Gurobi 13.0.1 solves. The mixed-integer programming (MIP) gap tolerance sets the certification criterion. The solver budgets are $t_{\mathrm{lim}}^{\mathrm{op}}$ for online control and $t_{\mathrm{lim}}^{\mathrm{ref}}$ for reference solves. For the initial-state distributions, $\mathcal N(\mu,\sigma^2)$ denotes a normal distribution with mean $\mu$ and variance $\sigma^2$.
Requested pack power follows from the WLTC Class~3b speed trace~\cite{UN_GTR15} and a backward-facing vehicle model~\cite{guzzella2013vehicle}. The simulation and postprocessing code, cell data, demand preprocessing, and configuration constants are archived in~\cite{c2CodeArchive}.
Nominal evaluation parameters are summarized in Table~\ref{tab:evaluation_protocol}.

\begin{table}[!htb]
\centering\scriptsize
\caption{Nominal Evaluation Parameters.}\label{tab:evaluation_protocol}
\setlength{\tabcolsep}{1.2pt}
\renewcommand{\arraystretch}{1.15}
\begin{tabular}{@{}lclc@{}}
\toprule
Quantity & Value & Quantity & Value\\\midrule
$N$ [-] & 20 & $N_p$ [-] & 10\\
$\Delta t$ [s] & 1 & Duration [s] & 1800\\
Demand & WLTC Class 3b & Solver threads [-] & 4\\
$t_{\mathrm{lim}}^{\mathrm{op}}$ [s] & 1 & $t_{\mathrm{lim}}^{\mathrm{ref}}$ [s] & 10\\
MIP-gap tolerance [-] & $10^{-4}$ & $\mathrm{SOH}^{\mathrm{EOL}}$ [-] & 0.80\\
$[v_{\mathrm{cell}}^{\min},v_{\mathrm{cell}}^{\max}]$ [V] & $[2,3.6]$ & $[v_{\mathrm{pack}}^{\min},v_{\mathrm{pack}}^{\max}]$ [V] & $[44.8,72]$\\
$i_{\mathrm{pack}}$ [A] & $[-3,18]$ & $\mathrm{SOC}_i$ [-] & $[0.05,1]$\\
$T_i$ [$^\circ$C] & $[11,44]$ & $\delta_P$ [-] & 0.01\\
$\mathrm{SOC}_i(0)$ [-] & $\mathcal N(0.8,0.001^2)$ & $\mathrm{SOH}_i(0)$ [-] & $\mathcal N(0.98,0.01^2)$\\
$i_{\mathrm{RC},i}(0)$ [A] & 0 & $T_i(0)$ [$^\circ$C] & 25\\
\multicolumn{4}{@{}l}{$(w_{\mathrm{SOH}},w_{\mathrm{SOC}},w_\lambda)=(10,4,1000)$ [-]}\\
\multicolumn{4}{@{}l}{$(\gamma_v,\gamma_{\mathrm{SOC}},\gamma_{\mathrm{SOH}})=(0.20,0.30,0.50)$ [-]}\\
\bottomrule
\end{tabular}
\end{table}

\subsection{Solver Performance Metrics}
An active step is a sampling instant with nonzero requested power. Thus, the active-step set is $\mathcal K_{\mathrm{act}}=\{k:p_{\mathrm{req}}(k)\neq0\}$, with count $K_{\mathrm{act}}=|\mathcal K_{\mathrm{act}}|$. For a scalar per-step metric $X$, summaries over the specified set are the mean $\bar X$, the percentile $X_\alpha$ at level $\alpha\in[0,100]$, and the maximum $X_{\max}$. The subscript 50 identifies the median.

Solver time $t(k)$ and branch-and-bound node count $n^{\mathrm{node}}(k)$ describe computational effort. We report $t_{50}$, $t_{95}$, $t_{\max}$ and the corresponding node-count summaries over $\mathcal K_{\mathrm{act}}$. The root-node percentage $\Phi_{\mathrm{root}}$ is the percentage of these steps with $n^{\mathrm{node}}\leq1$, including timeouts. The certification rate $\Phi_{\mathrm{cert}}$ is the percentage meeting the stated MIP-gap tolerance for the formulation being solved: full-subset MICP for the baseline and prefix-restricted MICP for the proposed controller. Speedup is the ratio of the baseline controller's solver time to the proposed controller's solver time at the same percentile, or at their respective maxima.

\subsection{Same-State Comparison and Restriction Metrics}
We isolate restriction cost by solving the two formulations with the same pre-control state, power horizon, and terminal-bound memory. The prefix solve uses $t_{\mathrm{lim}}^{\mathrm{op}}$, and the full-subset reference uses $t_{\mathrm{lim}}^{\mathrm{ref}}$. We draw states from two trajectories: one generated by the proposed controller and one generated exclusively by an independent cyclic engagement policy. Reference solves are evaluated offline along each trajectory. The cyclic-policy states provide a second distribution on which to assess the objective penalty.

We retain a state in the paired set $\mathcal K_{\mathrm{pair}}\subseteq\mathcal K_{\mathrm{act}}$ when both solves are certified, their objectives and returned slacks are finite, each slack is at most the detection threshold $\varepsilon_\lambda=0.01$ A, and the reference objective is nonzero. The number of retained states is $K_{\mathrm{pair}}=|\mathcal K_{\mathrm{pair}}|$; their percentage of active states is $\Phi_{\mathrm{pair}}=100K_{\mathrm{pair}}/K_{\mathrm{act}}$. For $k\in\mathcal K_{\mathrm{pair}}$, define the signed relative objective gap $g_{\mathrm{rel}}(k)$ by
\begin{equation}
g_{\mathrm{rel}}(k)=100\frac{J_{\mathrm{pref}}^\star(k)-J_{\mathrm{full}}^\star(k)}{J_{\mathrm{full}}^\star(k)}\quad[\%].
\end{equation}
Its mean, median, upper percentile, and maximum are $\bar g_{\mathrm{rel}}$, $g_{\mathrm{rel},50}$, $g_{\mathrm{rel},95}$, and $g_{\mathrm{rel},\max}$.

For the component index $c\in\{\mathrm{SOH},\mathrm{SOC},\lambda\}$, let $w_c$ be the corresponding objective weight in~\eqref{eq:ContrOBJ}. The formulation index $a\in\{\mathrm{full},\mathrm{pref}\}$ identifies full-subset or prefix optimization. Write $J_c^{a,\star}(\ell;k)$ for component $c$ at stage $\ell$ of the certified solution starting at sample $k$. Its weighted horizon average is $C_c^{a,\star}(k)$:
\begin{equation}
C_c^{a,\star}(k)=\frac{1}{N_p}\sum_{\ell\in\mathcal H_k}w_cJ_c^{a,\star}(\ell;k).
\end{equation}
The mean total difference $\Delta\bar J_{\mathrm{obj}}^{\mathrm{SS}}$ and weighted component difference $\Delta\bar J_{c,w}^{\mathrm{SS}}$ are
\begin{equation}
\begin{aligned}
\Delta\bar J_{\mathrm{obj}}^{\mathrm{SS}}&=\frac{1}{K_{\mathrm{pair}}}\sum_{k\in\mathcal K_{\mathrm{pair}}}\left[J_{\mathrm{pref}}^\star(k)-J_{\mathrm{full}}^\star(k)\right],\\
\Delta\bar J_{c,w}^{\mathrm{SS}}&=\frac{1}{K_{\mathrm{pair}}}\sum_{k\in\mathcal K_{\mathrm{pair}}}\left[C_c^{\mathrm{pref},\star}(k)-C_c^{\mathrm{full},\star}(k)\right].
\end{aligned}
\end{equation}
Superscript $\mathrm{SS}$ denotes same-state comparison; subscripts $\mathrm{obj}$ and $w$ identify the total objective and weighted components. Positive differences indicate an increased cost under prefix restriction.

\subsection{Closed-Loop Performance Metrics}
Power tracking, balancing, and cell usage describe the applied control actions. At an active step, relative absolute power-tracking error $e(k)$ and fraction of engaged cells $\phi_{\mathrm{eng}}(k)$ are
\begin{equation}
\begin{gathered}e(k)=100\frac{|p_{\mathrm{pack}}(k)-p_{\mathrm{req}}(k)|}{|p_{\mathrm{req}}(k)|}\quad[\%],\\
\phi_{\mathrm{eng}}(k)=\frac{m(k)}{N},\end{gathered}
\end{equation}
where $p_{\mathrm{pack}}(k)$ is the simulated plant power. We report $e_{50}$, $e_{99}$, $e_{\max}$, and $\bar\phi_{\mathrm{eng}}$ over $\mathcal K_{\mathrm{act}}$. Curtailment rate $\Phi_{\mathrm{curt}}$ is the percentage of active steps with returned slack exceeding $\varepsilon_\lambda$.

At each post-initial simulation state, cell-to-cell SOC dispersion is the population standard deviation
\begin{equation}
\sigma_{\mathrm{SOC}}(k)=100\sqrt{\frac{1}{N}\sum_{i\in\mathcal C}\left(\mathrm{SOC}_i(k)-\bar{\mathrm{SOC}}(k)\right)^2}\quad[\mathrm{pp}],
\end{equation}
where pp denotes percentage points. 

Switching count $n_{\mathrm{sw}}(k)$ measures individual cell transitions:
\begin{equation}
n_{\mathrm{sw}}(k)=\sum_{i\in\mathcal C}|S_i(k)-S_i(k-1)|.
\end{equation}
The mean $\bar n_{\mathrm{sw}}$ and 95th percentile $n_{\mathrm{sw},95}$ summarize cell transitions over active steps, excluding the first command.

For objective summaries, $\mathcal K_{\mathrm{usable}}$ contains the active steps at which the solver returns a usable horizon solution. The superscript $\mathrm{ret}$ denotes that returned solution, and $J_c^{\mathrm{ret}}(\ell;k)$ is its component-$c$ stage cost at prediction stage $\ell$ for sample $k$. Define the mean weighted component $\bar J_{c,w}^{\mathrm{CL}}$ and mean total objective $\bar J_{\mathrm{obj}}^{\mathrm{CL}}$ as
\begin{equation}
\begin{aligned}
\bar J_{c,w}^{\mathrm{CL}}&=\frac{1}{|\mathcal K_{\mathrm{usable}}|}\sum_{k\in\mathcal K_{\mathrm{usable}}}\frac{1}{N_p}\sum_{\ell\in\mathcal H_k}w_cJ_c^{\mathrm{ret}}(\ell;k),\\
\bar J_{\mathrm{obj}}^{\mathrm{CL}}&=\sum_{c\in\{\mathrm{SOH},\mathrm{SOC},\lambda\}}\bar J_{c,w}^{\mathrm{CL}}.
\end{aligned}
\end{equation}
The superscript $\mathrm{CL}$ identifies closed-loop operation. Each mean averages predicted horizon costs along the trajectory of the controller being evaluated, including usable incumbents at uncertified steps. 

\section{Results and Discussion}
\label{sec:results}
The evaluation compares solver time and certification, objective cost at identical states, and closed-loop control performance.

\subsection{Solver Time and Certification}
\begin{figure}[!htb]
\centering\includegraphics[width=\linewidth]{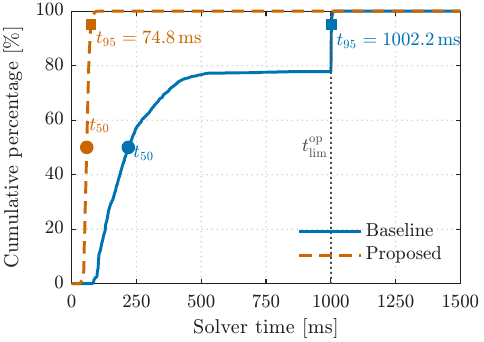}
\caption{Empirical solver-time distributions. The dotted vertical line marks $t_{\mathrm{lim}}^{\mathrm{op}}=1$ s; circles and squares identify $t_{50}$ and $t_{95}$, respectively.}\label{fig:runtime}
\end{figure}

Table~\ref{tab:runtime} and Figure~\ref{fig:runtime} summarize solver times, search effort, and solver-time distributions for the baseline and proposed controllers over the WLTC.

\begin{table}[!htb]
\centering\footnotesize
\caption{Solver-Time and Branch-and-Bound Statistics for the Nominal WLTC Cycle.}\label{tab:runtime}
\setlength{\tabcolsep}{1.3pt}
\renewcommand{\arraystretch}{1.15}
\begin{tabular}{lccccccc}\hline\hline
Controller & \shortstack{$t_{50}$\\{}[ms]} & \shortstack{$t_{95}$\\{}[ms]} & \shortstack{$t_{\max}$\\{}[ms]} & \shortstack{$\Phi_{\mathrm{root}}$\\{}[\%]} & \shortstack{$n^{\mathrm{node}}_{50}$\\{}[-]} & \shortstack{$n^{\mathrm{node}}_{95}$\\{}[-]} & \shortstack{$n^{\mathrm{node}}_{\max}$\\{}[-]}\\\hline
Baseline & 218.9 & 1002.2 & 1016.3 & 79.11 & 1 & 736 & 1763 \\
Proposed & 58.6 & 74.8 & 105.3 & 100.00 & 1 & 1 & 1 \\
\hline Speedup [-] & $3.7\times$ & $13.4\times$ & $9.7\times$ & -- & -- & -- & --\\\hline\hline\end{tabular}\end{table}

The baseline reaches 736 branch-and-bound nodes at the 95th percentile and 1763 in the worst case; every proposed-controller solve terminates at the root.

The largest runtime reduction occurs in the upper tail. The proposed controller reduces $t_{95}$ from 1002.2 to 74.8 ms, a factor of $13.4$, while $t_{50}$ decreases from 218.9 to 58.6 ms, a factor of $3.7$. The proposed controller's maximum solver time is 105.3 ms. Since both median times are below $t_{\mathrm{lim}}^{\mathrm{op}}$, reducing the upper tail matters more for real-time operation. Figure~\ref{fig:runtime} shows the narrower solver-time distribution of the proposed controller.

The baseline fails to certify optimality on 22.17\% of the active steps, whereas the proposed controller certifies all active solves. Both controllers return usable feasible actions throughout the cycle. Every baseline timeout returns a usable feasible incumbent, with a median MIP gap of 1.239\%. The baseline controller's principal computational limitation is therefore certifying optimality within $t_{\mathrm{lim}}^{\mathrm{op}}$; feasible control actions remain available at every step.

\subsection{Same-State Restriction Cost}
To quantify the objective cost of excluding nonprefix cell selections, we solve both controller problems from identical pre-control states. Table~\ref{tab:paired} compares their objective values along the proposed-controller trajectory and the independent cyclic-policy trajectory.

\begin{table}[!htb]
\centering\small
\caption{Same-State Restriction Cost on Jointly Certified, Uncurtailed Steps $\mathcal K_{\mathrm{pair}}$.}\label{tab:paired}
\setlength{\tabcolsep}{3pt}
\renewcommand{\arraystretch}{1.15}
\begin{tabular}{lcc}\hline\hline
Metric & \shortstack{Proposed-controller\\trajectory} & \shortstack{Independent\\trajectory}\\\hline
$K_{\mathrm{pair}}$ [-] (of 1800) & 1603 & 1571 \\
$\Phi_{\mathrm{pair}}$ [\%] & 89.06 & 87.28 \\
$\Delta\bar J_{\mathrm{obj}}^{\mathrm{SS}}$ [-] & +0.01464 & +0.00789 \\
$\Delta\bar J_{\mathrm{SOH},w}^{\mathrm{SS}}$ [-] & +0.01496 & +0.00807 \\
$\Delta\bar J_{\mathrm{SOC},w}^{\mathrm{SS}}$ [-] & -0.00032 & -0.00018 \\
$\bar g_{\mathrm{rel}}$ [\%] & 0.2552 & 0.1379 \\
$g_{\mathrm{rel},50}$ [\%] & 0.2484 & 0.1376 \\
$g_{\mathrm{rel},95}$ [\%] & 0.6295 & 0.3470 \\
$g_{\mathrm{rel},\max}$ [\%] & 0.8311 & 0.4672 \\
\hline\hline\end{tabular}\end{table}

On jointly certified, uncurtailed states, the proposed controller increases the mean objective on both evaluated trajectories. The mean relative gap is 0.2552\% along the proposed-controller trajectory and 0.1379\% along the independent trajectory, with positive medians of 0.2484\% and 0.1376\%, respectively. These medians show that the cost extends beyond isolated extreme states. The maximum gaps are 0.8311\% on the proposed-controller trajectory and 0.4672\% on the independent trajectory. The component differences show a health--balancing tradeoff: prefixes increase the SOH-weighted cost while partly offsetting that increase through a lower SOC-weighted cost. Neither controller curtails power on the paired states.

With the longer reference budget $t_{\mathrm{lim}}^{\mathrm{ref}}=10$ s, the baseline controller remains uncertified at 197 of the 1800 active states on the proposed-controller trajectory and 229 of the 1800 active states on the independent trajectory. The restriction-gap statistics therefore apply only to the jointly certified states in $\mathcal K_{\mathrm{pair}}$, which cover 89.06\% and 87.28\% of the respective active states.

\subsection{Closed-Loop Control Performance}
Table~\ref{tab:performance} compares the closed-loop performance of the baseline and proposed controllers under the operational solver limit.

\begin{table}[!htb]
\centering\small
\caption{Closed-Loop Performance over the Nominal WLTC Cycle.}\label{tab:performance}
\setlength{\tabcolsep}{4pt}
\renewcommand{\arraystretch}{1.15}
\begin{tabular}{lcc}\hline\hline
Metric & Baseline & Proposed\\\hline
\multicolumn{3}{l}{\textit{Closed-loop online objective}}\\
$\bar J_{\mathrm{obj}}^{\mathrm{CL}}$ [-] & 5.8505 & 5.7720 \\
$\bar J_{\mathrm{SOH},w}^{\mathrm{CL}}$ [-] & 5.7633 & 5.7705 \\
$\bar J_{\mathrm{SOC},w}^{\mathrm{CL}}$ [-] & 0.0848 & 0.0014 \\
$\bar J_{\lambda,w}^{\mathrm{CL}}$ [-] & 0.0025 & 0.0000 \\
\hline
\multicolumn{3}{l}{\textit{Curtailment}}\\
$\Phi_{\mathrm{curt}}$ [\%] & 0.1667 & 0.0000 \\
\hline
\multicolumn{3}{l}{\textit{Cell-to-cell SOC regulation}}\\
$\bar\sigma_{\mathrm{SOC}}$ [pp] & 0.8150 & 0.0932 \\
$\sigma_{\mathrm{SOC},95}$ [pp] & 1.2633 & 0.1433 \\
$\sigma_{\mathrm{SOC},\max}$ [pp] & 1.3720 & 0.2304 \\
\hline
\multicolumn{3}{l}{\textit{Power-tracking error}}\\
$e_{50}$ [\%] & 0.0154 & 0.0157 \\
$e_{99}$ [\%] & 0.0290 & 0.0289 \\
$e_{\max}$ [\%] & 0.9826 & 0.0421 \\
\hline
\multicolumn{3}{l}{\textit{Cell usage and switching}}\\
$\bar\phi_{\mathrm{eng}}$ [-] & 0.7074 & 0.7068 \\
$\bar n_{\mathrm{sw}}$ [-] & 1.8410 & 6.3513 \\
$n_{\mathrm{sw},95}$ [-] & 10 & 12 \\
\hline\hline\end{tabular}\end{table}

The proposed controller reduces the cycle-mean, 95th-percentile, and maximum cell-to-cell SOC dispersions by factors of 8.7, 8.8, and 6.0, respectively. The SOH-weighted objective contribution increases by approximately 0.125\%, so tighter balancing accompanies only a small change in aggregate SOH-related cost. Averaged along each controller's trajectory and including uncertified baseline incumbents, the proposed controller has a lower mean returned horizon objective, with lower SOC and curtailment contributions.

\begin{figure}[!htb]
\centering\includegraphics[width=\linewidth]{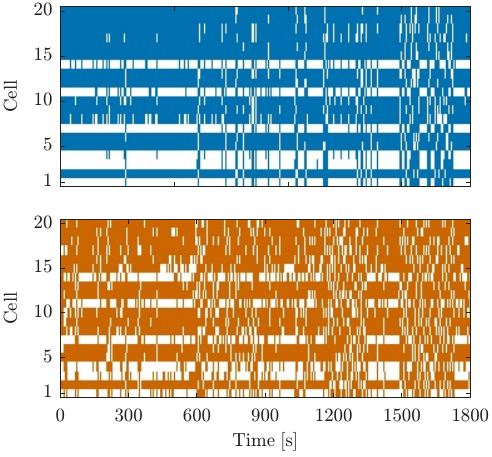}
\caption{Cell engagement over the WLTC: baseline controller (top, blue) and proposed controller (bottom, orange). Colored entries denote engagement; white entries denote bypass.}\label{fig:engagement}
\end{figure}

Both controllers engage about 70.7\% of the cells on average. The balancing improvement therefore accompanies a redistribution of engagement among cells, with little change in their average number. Figure~\ref{fig:engagement} shows this redistribution as the ranking evolves.

Both controllers have median tracking errors near 0.016\% and 99th-percentile errors near 0.029\%. The baseline curtails power on 0.1667\% of active steps and reaches a maximum error of 0.9826\% at a curtailed step. The proposed controller has no detected curtailment and a maximum error of 0.0421\%. Tracking therefore improves primarily through the suppression of rare large deviations.

The proposed controller increases mean cell transitions per active step by a factor of 3.45, from 1.8410 to 6.3513, while the 95th percentile rises from 10 to 12. The larger relative increase in the mean indicates more frequent routine reconfiguration, with a smaller increase in the upper switching quantile.

\section{Conclusion}
\label{sec:conclusion}
We propose a ranking-based feasible-set restriction for real-time MI-MPC and apply it to cell-bypass reconfigurable battery packs. Fixed weights on normalized SOH, SOC, and voltage headroom rank the cells, and a monotone prefix constraint reduces per-stage cell selection to a choice of engagement count while leaving the model, objective, and physical constraints unchanged. The resulting certificates apply to the ranked-prefix formulation, not to the excluded nonprefix subsets. The formulation adds only linear inequalities, remains compatible with standard mixed-integer convex solvers, and requires no offline training, custom branching, or solver modification.

In the nominal 20-cell WLTC case with a 1 s solver limit, the restriction increases the certified-optimality rate from 77.83\% to 100\%, reduces the 95th-percentile solver time by a factor of 13.4, and decreases the cycle-mean cell-to-cell SOC standard deviation by a factor of 8.7 without power curtailment. Same-state comparisons on two trajectories yield mean restriction gaps of 0.26\% and 0.14\% on jointly certified, uncurtailed steps, while mean cell switching increases by a factor of 3.45. Future work should address evaluation across operating conditions and controller settings, comparison with alternative cell-selection methods, switching penalties or dwell-time constraints, and hardware tests of switching cost, efficiency, and safe full-load commutation.

\sloppy
\bibliographystyle{IEEEtran}
\bibliography{IEEEabrv,references}

@STRING{IEEE_J_TTE        = "{IEEE} Trans. Transport. Electrific."}

@STRING{IEEE_J_AC         = "{IEEE} Trans. Autom. Control"}

@STRING{IEEE_J_CST        = "{IEEE} Trans. Control Syst. Technol."}

@STRING{IEEE_J_IE         = "{IEEE} Trans. Ind. Electron."}

@STRING{IEEE_O_ACC        = "{IEEE} Access"}

@STRING{IEEE_M_IE         = "{IEEE} Ind. Electron. Mag."}

@book{rawlings2020model,
  author    = {Rawlings, James B. and Mayne, David Q. and Diehl, Moritz M.},
  title     = {Model Predictive Control: Theory, Computation, and Design},
  edition   = {2nd},
  publisher = {Nob Hill Publishing},
  year      = {2017},
  url       = {https://sites.engineering.ucsb.edu/~jbraw/mpc/}
}

@article{bemporad1999control,
  author  = {Bemporad, Alberto and Morari, Manfred},
  title   = {Control of Systems Integrating Logic, Dynamics, and Constraints},
  journal = {Automatica},
  volume  = {35},
  pages   = {407--427},
  year    = {1999}
}

@inproceedings{richards2005mixed,
  author    = {Richards, Arthur and How, Jonathan P.},
  title     = {Mixed-Integer Programming for Control},
  booktitle = {Proc. Amer. Control Conf. ({ACC})},
  pages     = {2676--2683},
  year      = {2005}
}

@article{marcucci2020warm,
  author  = {Marcucci, Tobia and Tedrake, Russ},
  title   = {Warm Start of Mixed-Integer Programs for Model Predictive Control of Hybrid Systems},
  journal = IEEE_J_AC,
  volume  = {66},
  number  = {6},
  pages   = {2433--2448},
  year    = {2020}
}

@inproceedings{hespanhol2019structure,
  author    = {Hespanhol, Pedro and Quirynen, Rien and {Di Cairano}, Stefano},
  title     = {A Structure Exploiting Branch-and-Bound Algorithm for Mixed-Integer Model Predictive Control},
  booktitle = {Proc. 18th Eur. Control Conf. ({ECC})},
  year      = {2019},
  pages     = {2763--2768}
}

@article{han2020next,
  author  = {Han, Weiji and Wik, Torsten and Kersten, Anton and Dong, Guangzhong and Zou, Changfu},
  title   = {Next-Generation Battery Management Systems: Dynamic Reconfiguration},
  journal = IEEE_M_IE,
  volume  = {14},
  number  = {4},
  pages   = {20--31},
  year    = {2020}
}

@article{muhammad2019reconfigurable,
  author  = {Muhammad, Shaheer and Rafique, M. Usman and Li, Shuai and Shao, Zili and Wang, Qixin and Liu, Xue},
  title   = {Reconfigurable Battery Systems: A Survey on Hardware Architecture and Research Challenges},
  journal = {{ACM} Trans. Des. Autom. Electron. Syst.},
  volume  = {24},
  number  = {2},
  pages   = {1--27},
  year    = {2019}
}

@article{ci2016reconfigurable,
  author  = {Ci, Song and Lin, Ni and Wu, Dalei},
  title   = {Reconfigurable Battery Techniques and Systems: A Survey},
  journal = IEEE_O_ACC,
  volume  = {4},
  pages   = {1175--1189},
  year    = {2016}
}

@article{farakhor2022novel,
  author  = {Farakhor, Amir and Wu, Di and Wang, Yebin and Fang, Huazhen},
  title   = {A Novel Modular, Reconfigurable Battery Energy Storage System: Design, Control, and Experimentation},
  journal = IEEE_J_TTE,
  volume  = {9},
  number  = {2},
  pages   = {2878--2890},
  year    = {2023}
}

@article{farakhor2023scalable,
  author  = {Farakhor, Amir and Wu, Di and Wang, Yebin and Fang, Huazhen},
  title   = {Scalable Optimal Power Management for Large-Scale Battery Energy Storage Systems},
  journal = IEEE_J_TTE,
  volume  = {10},
  number  = {3},
  pages   = {5002--5016},
  year    = {2024}
}

@article{mondoha2021nonlinear,
  author  = {Mondoha, Abouzede and Sabatier, Jocelyn and Lanusse, Patrick and Tippmann, Simon and Farges, Christophe},
  title   = {Nonlinear Model Predictive Control for a Simulated Reconfigurable Battery Pack},
  journal = {IFAC-PapersOnLine},
  volume  = {54},
  number  = {6},
  pages   = {353--358},
  year    = {2021}
}

@article{pinter2025comparative,
  author  = {Pinter, Zoltan Mark and Rohde, Gunnar and Marinelli, Mattia},
  title   = {Comparative Analysis of Rule-Based and Model-Predictive Control Algorithms in Reconfigurable Battery Systems for {EV} Fast-Charging Stations},
  journal = {J. Energy Storage},
  volume  = {116},
  pages   = {116008},
  year    = {2025}
}

@article{ebrahimi2026safe,
  author  = {Ebrahimi, Iman and de Castro, Ricardo},
  title   = {Design and Experimental Validation of a Safe Control Strategy for Reconfigurable Batteries},
  journal = IEEE_J_CST,
  volume  = {34},
  number  = {3},
  pages   = {1404--1418},
  year    = {2026}
}

@article{wei2026reliable,
  author  = {Wei, Meng and Yin, Jingyuan and Xu, Guoning and Li, Zhuohan and Peng, Jinkai and Huo, Qunhai and Wei, Tongzhen},
  title   = {A Reliable Dynamically Reconfigurable Battery Topology Family and Hybrid Control Strategy Suitable for Multiple Operating Conditions},
  journal = IEEE_J_IE,
  volume  = {73},
  number  = {6},
  pages   = {8555--8566},
  year    = {2026}
}

@article{liu2026learning,
  author  = {Liu, Pei and Li, Da and Zhu, Jin and Cheng, Boyuan and Wu, Yu and Qi, Yang and Li, Weilin},
  title   = {A Learning-Based Method for Battery Health Balancing and Capacity Maximization in Reconfigurable Battery Systems},
  journal = {J. Energy Storage},
  volume  = {166},
  pages   = {122406},
  year    = {2026}
}

@article{irshayyid2026realtime,
  author  = {Irshayyid, Ali and Yang, Wanqun and Chen, Jun},
  title   = {Real-Time Balancing Control of Reconfigurable Battery Packs Using Reinforcement Learning},
  journal = IEEE_J_TTE,
  year    = {2026},
  note    = {{Early Access}; published online Apr. 10, 2026.}
}

@article{geng2025fundamental,
  author  = {Geng, Changyou and Ren, Dezhi and Mao, Enkai and Zou, Changfu and Va{\v{s}}ak, Mario and Zheng, Xinyi and Han, Weiji},
  title   = {Fundamental Techniques for Optimal Control of Reconfigurable Battery Systems: System Modeling and Feasible Search Space Construction},
  journal = {arXiv preprint arXiv:2501.03028},
  year    = {2025}
}

@article{geng2026unified,
  author  = {Geng, Changyou and Ren, Dezhi and Mao, Enkai and Zheng, Xinyi and Va{\v{s}}ak, Mario and Zou, Changfu and Han, Weiji},
  title   = {A Unified Modeling Framework of Reconfigurable Battery Systems for Optimal Control},
  journal = IEEE_J_TTE,
  volume  = {12},
  number  = {3},
  pages   = {5336--5347},
  year    = {2026}
}

@inproceedings{vskegro2023analysis,
  author    = {{\v{S}}kegro, Albert and Zou, Changfu and Wik, Torsten},
  title     = {Analysis of Potential Lifetime Extension Through Dynamic Battery Reconfiguration},
  booktitle = {Proc. 25th Eur. Conf. Power Electron. Appl. ({EPE}'23 {ECCE} Europe)},
  pages     = {1--11},
  year      = {2023}
}

@article{ouyang2025mathematical,
  author  = {Ouyang, Quan and {\v{S}}kegro, Albert and Hu, Lin and Han, Weiji and Ren, Dezhi and Geng, Changyou and Wik, Torsten and Zou, Changfu},
  title   = {Mathematical Modeling for Reconfigurable Battery Systems With Parallel--Series Connections},
  journal = IEEE_J_CST,
  volume  = {34},
  number  = {1},
  pages   = {186--202},
  year    = {2026}
}

@article{allafi_lumped_2018,
  author  = {Allafi, Walid and Zhang, Cheng and Uddin, Kotub and Worwood, Daniel and Dinh, Truong Quang and Ormeno, Pedro Ascencio and Li, Kang and Marco, James},
  title   = {A Lumped Thermal Model of {Lithium-Ion} Battery Cells Considering Radiative Heat Transfer},
  journal = {Appl. Therm. Eng.},
  volume  = {143},
  pages   = {472--481},
  year    = {2018}
}

@book{williams2013model,
  author    = {Williams, H. Paul},
  title     = {Model Building in Mathematical Programming},
  edition   = {5th},
  publisher = {Wiley},
  year      = {2013}
}

@article{vskegro2026system,
  title={System-level assessment of dynamic reconfiguration for lifetime and cost outcomes in electric vehicle battery packs},
  author={{\v{S}}kegro, Albert and Wik, Torsten and Bijlenga, Bo and Bessman, Alexander and Zou, Changfu},
  journal={Nat. Commun.},
  volume={17},
  number={1},
  pages={5980},
  year={2026},
  doi={10.1038/s41467-026-74951-8}
}

@misc{UN_GTR15,
  author      = {{United Nations}},
  title       = {{UN Global Technical Regulation No. 15: Worldwide
                  Harmonized Light Vehicles Test Procedure (WLTP),
                  Amendment 6}},
  institution = {United Nations Economic Commission for Europe},
  year        = {2021},
  url         = {https://unece.org/sites/default/files/2021-01/ECE-TRANS-180a15am6e.pdf}
}

@book{guzzella2013vehicle,
  author    = {Guzzella, Lino and Sciarretta, Antonio},
  title     = {Vehicle Propulsion Systems: Introduction to Modeling
               and Optimization},
  edition   = {3},
  publisher = {Springer},
  year      = {2013}
}

@misc{c2CodeArchive,
  author = {Albert {\v{S}}kegro},
  title = {Simulation and post-processing code for `{R}educing Combinatorial Redundancy in Mixed-Integer {MPC} via Ranking-Based Feasible-Set Restriction for Reconfigurable Batteries'},
  howpublished = {Zenodo},
  year = {2026},
  doi = {10.5281/zenodo.22885592},
  url = {https://doi.org/10.5281/zenodo.22885592}
}

@article{ma2025datadriven,
  author={Ma, Xutao and Ning, Chao and Du, Wenli and Shi, Yang},
  title={Data-Driven Distributionally Robust Mixed-Integer Control Through Lifted Control Policy},
  journal={{IEEE} Trans. Autom. Control},
  volume={70}, number={9}, pages={6268--6275}, year={2025},
  doi={10.1109/TAC.2025.3558138}
}

@incollection{margot2009symmetry,
  author={Margot, Fran{\c{c}}ois},
  title={Symmetry in Integer Linear Programming},
  booktitle={50 Years of Integer Programming 1958--2008},
  publisher={Springer}, pages={647--686}, year={2010},
  doi={10.1007/978-3-540-68279-0_17}
}

@IEEEtranBSTCTL{BSTcontrol,
  CTLdash_repeated_names = {no}
}

@article{knueven2018,
 author={Knueven, Ben and Ostrowski, Jim and Watson, Jean-Paul},
 title={Exploiting Identical Generators in Unit Commitment},
 journal={{IEEE} Trans. Power Syst.}, volume={33}, number={4}, pages={4496--4507}, year={2018}, doi={10.1109/TPWRS.2017.2783850}
}

\end{document}